\documentclass[preprint2]{aastex}

\usepackage{upgreek}
\usepackage{lineno}
\usepackage{color}

\newcommand{\g}{$\gamma$}
\newcommand{\F}{\textit{Fermi}}

\slugcomment{Manuscript prepared for ApJL}

\shorttitle{The variable \g-ray pulsar PSR~J2021+4026}
\shortauthors{The \F~LAT collaboration}

\begin{document}

\title{PSR~J2021+4026 in the Gamma Cygni region: the first variable \g-ray
pulsar seen by 
the \F~LAT}

\author{
A.~Allafort\altaffilmark{1}, 
L.~Baldini\altaffilmark{2}, 
J.~Ballet\altaffilmark{3}, 
G.~Barbiellini\altaffilmark{4,5}, 
M.~G.~Baring\altaffilmark{6}, 
D.~Bastieri\altaffilmark{7,8}, 
R.~Bellazzini\altaffilmark{9}, 
E.~Bonamente\altaffilmark{10,11}, 
E.~Bottacini\altaffilmark{1}, 
T.~J.~Brandt\altaffilmark{12}
% ...
}
\altaffiltext{1}{W. W. Hansen Experimental Physics Laboratory, Kavli Institute
for Particle Astrophysics and Cosmology, Department of Physics and SLAC National
Accelerator Laboratory, Stanford University, Stanford, CA 94305, USA}
\altaffiltext{2}{Universit\`a  di Pisa and Istituto Nazionale di Fisica
Nucleare, Sezione di Pisa I-56127 Pisa, Italy}
\altaffiltext{3}{Laboratoire AIM, CEA-IRFU/CNRS/Universit\'e Paris Diderot,
Service d'Astrophysique, CEA Saclay, 91191 Gif sur Yvette, France}
\altaffiltext{4}{Istituto Nazionale di Fisica Nucleare, Sezione di Trieste,
I-34127 Trieste, Italy}
\altaffiltext{5}{Dipartimento di Fisica, Universit\`a di Trieste, I-34127
Trieste, Italy}
\altaffiltext{6}{Rice University, Department of Physics and Astronomy, MS-108,
P. O. Box 1892, Houston, TX 77251, USA}
\altaffiltext{7}{Istituto Nazionale di Fisica Nucleare, Sezione di Padova,
I-35131 Padova, Italy}
\altaffiltext{8}{Dipartimento di Fisica e Astronomia ``G. Galilei'',
Universit\`a di Padova, I-35131 Padova, Italy}
\altaffiltext{9}{Istituto Nazionale di Fisica Nucleare, Sezione di Pisa, I-56127
Pisa, Italy}
\altaffiltext{10}{Istituto Nazionale di Fisica Nucleare, Sezione di Perugia,
I-06123 Perugia, Italy}
\altaffiltext{11}{Dipartimento di Fisica, Universit\`a degli Studi di Perugia,
I-06123 Perugia, Italy}
\altaffiltext{12}{NASA Goddard Space Flight Center, Greenbelt, MD 20771, USA}
% ...
\begin{abstract}

Long-term monitoring of PSR~J2021+4026 in the heart of the Cygnus region with the \F~Large Area
Telescope (LAT) unveiled a sudden decrease in flux above
100~MeV over a
time scale shorter than a week. 
The ``jump'' was near MJD~55850 (2011 October 16), with the flux
decreasing from $(8.33\pm0.08) \times 10^{-10}$~erg
cm$^{-2}$ s$^{-1}$ to $(6.86\pm0.13) \times 10^{-10}$~erg cm$^{-2}$ s$^{-1}$.
Simultaneously,  the frequency spindown rate increased from $(7.8\pm 0.1)
\times 10^{-13}$~Hz s$^{-1}$ to $(8.1\pm 0.1) \times 10^{-13}$~Hz
s$^{-1}$. 
Significant ($>5\sigma$) changes in the pulse profile and marginal
($<3\sigma$) changes in the emission spectrum
occurred at the same time.
There is also evidence for a small, steady flux increase over the three years preceding MJD~55850. 
This is the first observation at \g-ray energies of mode changes and intermittent behavior,
observed at radio wavelengths for other pulsars. We argue that the change in
pulsed \g-ray emission is due to a change in emission beaming
and we speculate that it is precipitated by a shift in the
magnetic field
structure, leading to a change of either effective magnetic inclination or
effective current.

\end{abstract}


\section{Introduction}
Pulsars are the largest high-energy \g-ray source class in the Milky Way,
with 117 characterized in the second \F~Large Area Telescope (LAT) pulsar catalog (2PC,
\citealt{2PCpaper}).
Early \g-ray observations suggested glitch-associated
pulse and flux changes in the Crab \citep{greisen1975} and Vela \citep{grenier1988} pulsars, 
unconfirmed by additional observations \citep{nolan2003}.
Steady \g-ray fluxes and pulse profiles on timescales longer than
those needed for
pulsar detections has been an axiom \citep[e.g.,][hereafter 2FGL]{2FGLpaper}.

PSR J2021+4026 (hereafter J2021+4026) was discovered in a blind frequency search
using LAT data \citep{LATblindsearch}. Its spin frequency $f \sim 3.8$~Hz and
frequency derivative $\dot{f}\sim -8 \times 10^{-13}$~Hz s$^{-1}$ point to a young, energetic pulsar (characteristic age $\tau_{c}$=77 kyr, spindown
power $\dot{E}_\mathrm{SD}\sim 10^{35}$~erg s$^{-1}$). 
Radio and optical searches did not yield any plausible counterparts,
while deep observations with
\textit{Chandra} and \textit{XMM-Newton} led to an 
association with the X-ray source S20 \citep{weisskopf2011,trepl2010}, from
which X-ray pulsations were recently detected
\citep{lin2013}. J2021+4026 is seen within the radio shell of 
the supernova remnant (SNR) G78.2+2.1 \citep[e.g.,][]{ladouceur2008}, also an
extended \g-ray source with $\sim 0\fdg6$ radius \citep{LATextended}. 
A tentative association of these two sources implies a pulsar distance of
$\sim1.5$ kpc.

Using AGILE observations from 2007 November to 2009 August,
\citet{chen2011} reported variability in 1AGL~J2022+4032,
positionally coincident with J2021+4026, but concluded that it was
more
likely due to another source along the line of sight. This
Letter
reports \F~LAT observation of a discrete change in the \g-ray flux
and frequency derivative of
J2021+4026 and further results from its long-term monitoring.

\section{Observations and likelihood analysis}\label{sec:anal}
We analyzed $\sim$52 months of LAT data from 2008 August 4 to 2012 December 11.
We selected P7REP\_SOURCE ``photon'' class events \citep{bregeon2013,line2013}, 
in a 15\arcdeg~radius Region Of Interest (ROI), 
with energies from 100~MeV to 300~GeV and zenith angles $\leq 100\arcdeg$. 
We excluded time intervals when the LAT rocking angle was $> 52\arcdeg$ or
the zone defined by the zenith cut intersected the ROI.

We characterized the spectrum of J2021+4026 through a binned likelihood
fit over a $14\arcdeg \times 14\arcdeg$ region centered on the pulsar,
with  $0\fdg1$~angular grid. We used logarithmically-spaced
energy bins from 100 MeV to 300 GeV (16 below 10 GeV, 8 above).
The combined likelihood technique
\citep[e.g.,][]{ackermann2012} treats
photons converting in the front and back tracker sections separately, 
to exploit the former's higher angular resolution \citep{atwood2009}. 
We used the P7REP\_SOURCE\_V15 LAT Instrument Response Functions (IRFs) and associated
diffuse emission models\footnote{The P7REP data, IRFs, and
diffuse models (gll$\_$iem$\_$v05.fit, iso$\_$source$\_$front$\_$v05.txt,
iso$\_$source$\_$back$\_$v05.txt) will be available at
\url{http://fermi.gsfc.nasa.gov/ssc/}.} and LAT
\textit{Science Tools} v09r32p01. 
The isotropic background and the residual Earth limb emission were kept fixed.
The Galactic diffuse emission was multiplied by a power law with free
normalization and spectral index. Our model includes the known extended sources in
the region, i.e., the Cygnus-X cocoon
\citep{LATcocoon} and SNR~G78.2+2.1 \citep{LATextended}, not included in
previous analyses
(2FGL, 2PC). We also included 2PC pulsars, and additional 2FGL sources with high significance 
(average Test Statistic\footnote{The Test
Statistic for source detection from maximum likelihood ratio (see,
e.g., 2FGL).} $\mathrm{TS} > 100$), or within 4$\arcdeg$ of J2021+4026. 
Known flaring sources in the region
\citep{FAVApaper} are included in the model. 
Source spectra were modeled using the functional forms described in the catalogs with 
all the spectral parameters free. 
We then searched a TS map of the region for excesses, used as seeds 
to determine positions and spectra of new sources (see 2FGL). We
found five, modeled with log-parabola
spectra, at the epoch J2000 positions
$(\mathrm{R.A}.,\mathrm{Dec}) = (310\fdg52, 42\fdg06)$,
$(306\fdg66, 40\fdg05)$, $(309\fdg65, 42\fdg22)$,
$(308\fdg67, 43\fdg05)$, $(312\fdg44, 44\fdg38)$.

J2021+4026's position was set to \textit{Chandra} S20's. 
Its \g-ray spectrum was modeled using a power law with exponential cutoff (PLEC, Equation~\ref{eq:plec}), 
with $b$ free for the phase-averaged spectral analysis and $b=1$ (PLEC1) for the phase-resolved analysis 
in Section~\ref{sec:psrcar}.
\begin{equation}\label{eq:plec}
 \frac{\mathrm{d}N}{\mathrm{d}E}= N_0 \left(\frac{E}{E_0}\right)^{-\Gamma}
\exp\left[-\left(\frac{E}{E_\mathrm{c}}\right)^b\right]
\end{equation}
PLEC1 represents expectations for high-altitude magnetospheric
emission. 
Phase-averaged spectra are usually better fit with $b<1$, due to the
superposition of several PLEC1 components with different photon
indices and cutoff energies \citep{abdo2010}.

\section{Flux variability}

Following \citet{chen2011}, we searched for flux variability around
J2021+4026 at energies $E>100$~MeV and $E>1$~GeV, applying the
method in 2FGL. 
We first fit the data over the entire time range. 
Then, we divided the range into 7 and 30-day time bins and re-fit, 
allowing free normalizations for all sources and fixing the other spectral
parameters to their long-term average. The fit was then repeated in each time
bin by also fixing the source of interest's normalization to its long-term average. 
The Galactic diffuse normalization was fit in each time bin, 
verifying a posteriori its compatibility with a constant.
Following 2FGL Eq.~4, the fit maximum likelihood values established the probability $P$ that
the observed fluctuations are stochastic only.
A $2\%$ flux systematic error accounts for exposure uncertainties
between
different epochs.

We applied this procedure to J2021+4026, 
to SNR~G78.2+2.1 (seen in the same direction, but separable from
the pulsar at high energies due to its extension\footnote{
The LAT~68\% containment radius for front-converting events is $0\fdg7$
($< 0\fdg2$) at 1 (10)~GeV \citep{ackermann2012perf,bregeon2013}.}) and to
J2021+3651,
located $3\fdg5$ away, with spectrum and flux similar to J2021+4026's.
Both SNR~G78.2+2.1 and J2021+3651 show constant fluxes ($P >0.4$ in all
cases), while J2021+4026 shows significant variability:
$P_\mathrm{30\:days}=7\times 10^{-8}$ ($1\times 10^{-10}$),
$P_\mathrm{7\:days}=6\times 10^{-3}$ ($2\times10^{-4}$) at energies
$>100$~ MeV ($>1$~GeV).

\begin{figure*}
\epsscale{1.8}
\plotone{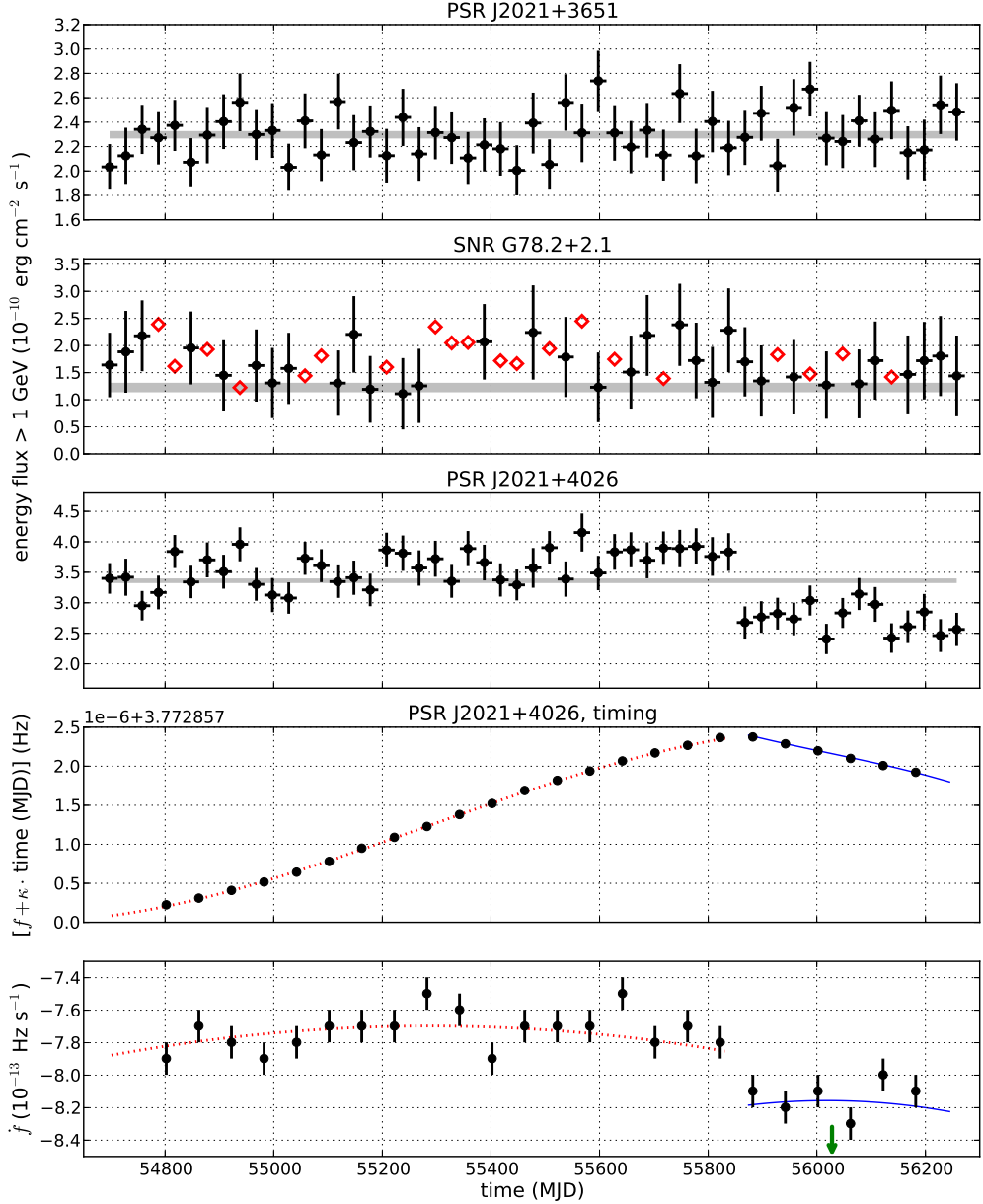}
\caption{Three top panels: energy flux ($E> 1$~GeV) versus time in
30-day bins for J2021+3651, SNR~G78.2+2.1, and J2021+4026. 
The gray bands show the average source fluxes for all data.
Statistical uncertainties only. We report 95\%
confidence level upper limits (red diamonds) for time bins where TS $< 4$. 
Two bottom panels (Section~\ref{sec:timing}):
for J2021+4026, 
$f+\kappa \cdot$~time (MJD), with frequency $f$ 
and $\kappa=6.9 \times 10^{-8}$~Hz day$^{-1}$, and frequency
derivative $\dot{f}$, versus time, from the periodicity search in 60-day
windows (points), 
and from the timing solutions for MJD~$<55850$ (red dotted line) and $>55850$
(blue solid line). The green arrow indicates the epoch of the X-ray
pulsation detection \citep{lin2013}.}\label{fig:1}
\end{figure*}

Figure~\ref{fig:1} shows the $>1$~GeV energy flux for the three sources in 30-day bins. J2021+4026 shows an abrupt $\sim 20\%$ flux decrease near MJD~55850 (Table~\ref{tbl-2}),
confirmed at energies $>100$~MeV and for 7-day bins. 
We exclude that this drop is due to systematic effects since there is no
analogous drop for the two other sources, observed simultaneously.
No significant changes in the \F\ observing strategy occurred near
MJD~55850.
We verified that fixing J2021+4026's normalization to its average
yields negative residuals consistent with a point-like source at the pulsar
position in all energy bands after MJD~55850.

\begin{deluxetable}{lcc}
\tabletypesize{\scriptsize}
\tablecaption{J2021+4026's properties
before and after the jump.\label{tbl-2}}
\tablewidth{0pt}
\tablehead{
\colhead{Time Range(MJD)} & \colhead{54682--55850} & \colhead{55850--56273}\\}
\startdata
number of days & 1167 & 423 \\
$F_{\gamma}$\tablenotemark{a} $>0.1$~GeV & $8.33\pm 0.08$ & $6.86\pm0.13$ \\
$F_{\gamma}$\tablenotemark{a} $>1$~GeV & $3.57\pm0.05$ & $2.74\pm0.06$ \\
$\dot{f}$\tablenotemark{b} &$-7.6978\pm0.0007$ &
$-8.166\pm0.002$ \\
$\delta_{P1}$\tablenotemark{c} &0.19$\pm$0.02&0.13$\pm$0.02 \\
$\Delta_{12}$\tablenotemark{d} &0.505$\pm$0.005&0.565$\pm$0.006 \\
$\delta_{P2}$\tablenotemark{c} &0.176$\pm$0.007&0.174$\pm$0.006 \\
$\Delta_{1BR}$\tablenotemark{d} &0.229$\pm$0.008& --\\
$\delta_{BR}$\tablenotemark{c} &0.11$\pm$0.02& -- \\
P1/P2\tablenotemark{e}& $0.54\pm0.06$ & $0.24\pm0.03$ \\
BR/P2\tablenotemark{e}&  $0.16\pm0.03$ & -- \\ 
constant/P2\tablenotemark{e}&  $1.83\pm0.14$ & $1.09\pm0.06$ \\ 
\enddata
\tablecomments{Statistical uncertainties only. {For details on parameters, see Sections}~\ref{sec:timing} and~\ref{sec:psrcar}.}
\tablenotetext{a}{$10^{-10}$ erg cm$^{-2}$ s$^{-1}$.}
\tablenotetext{b}{At the reference epoch for the two timing solutions,
$10^{-13}$ Hz s$^{-1}$.}
\tablenotetext{c}{Peak FWHM ($E>0.1$~GeV).}
\tablenotetext{d}{Phase lag between peaks ($E>0.1$~GeV).}
\tablenotetext{e}{Ratios of the peak amplitudes or
constant-level-to-P2
amplitude ($E>0.1$~GeV).}
\end{deluxetable}

Figure 1 also suggests a steady flux increase for J2021+4026
before the drop near MJD~55850.
A $\chi^2$ fit of a linear function of time versus flux in 30-day
bins\footnote{We
assumed a 2\% systematic flux uncertainty, as for the variability
test.} 
before MJD~55850 gives a $4\% \pm 2\%$~year$^{-1}$ flux increase, preferred
over the
constant flux hypothesis at the $\sim 3.3\sigma$ level
for both $>100$~MeV and $>1$~GeV. The $\chi^2$ test
applied to J2021+3651 favors a constant flux. 
We further assessed this trend independently of any functional dependency using
the Kendall rank correlation test. 
We obtain a Kendall coefficient $\tau=0.78$~$(0.71)$ for J2021+4026 for $>100$~MeV ($>1$~GeV): 
the probability\footnote{We take trial factors due to truncating
the sample at
MJD 55850 into account with a Monte Carlo simulation,  where we calculate the maximum
$\tau$ for stopping after 4 different 30-day bins around MJD~55850.
The 120-day time scale is independently
constrained by  the timing analysis 
in Section~\ref{sec:timing} (two 60-day bins in the periodicity
search).} of this coming from
stochastic fluctuations of a steady flux is 0.0005 (0.01), indicating a
monotonic increase with time at the $\sim 3.5 \sigma$ ($\sim 3.2 \sigma$) level. 
For J2021+3651, $\tau=0.11$~$(0.09)$: the same probability is 0.38 (0.41).

\section{Pulsar timing}\label{sec:timing}
\begin{figure*}
\epsscale{1.6}
\plotone{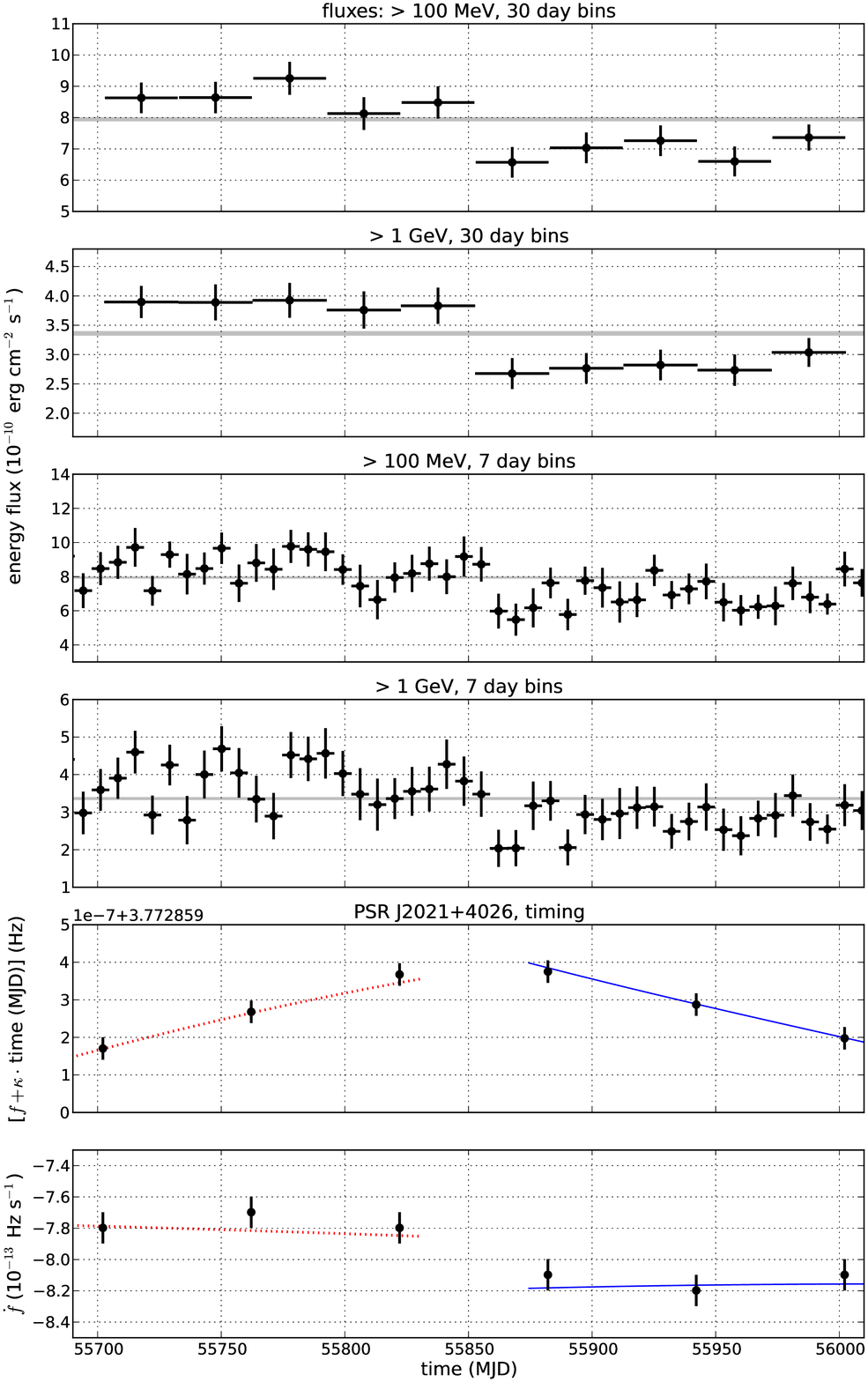}
\caption{Four top panels: energy flux versus time for J2021+4026 during $\sim 300$ days
centered at MJD~55850, in different time bins and energy bands. 
Two bottom panels: $f$ and $\dot{f}$
(further description in 
Figure~\ref{fig:1}).}\label{fig:2new}
\end{figure*}

To investigate the origin of the flux drop, we monitored the evolution of the pulsar timing parameters. 
We divided the entire time range into 60-day bins, where pulsations are clearly detectable, yet we
can neglect timing noise and approximate the frequency evolution
as a linear function,
\begin{equation}
 f(t)=f_0 + f_1\times (t-t_0)\, .
\end{equation}
For each bin, we used the Z$^{2}_{n}$ ($n=4$) test \citep{buccheri1983} to
search the $f_0$--$f_1$ space for periodicity ($f_0$ and $f_1$ represent the
frequency $f$ and frequency derivative $\dot{f}$, respectively, in each bin).
Figure~\ref{fig:1} shows that near MJD~55850
$\dot{f}$ suddenly decreases by $\sim 5\times 10^{-14}$ Hz s$^{-1}$, 
i.e. $\sim 4\%$ of the initial value. 
This is reflected as a change of the frequency evolution slope, 
while $f$ does not change appreciably. 
The $\dot{f}$ change is simultaneous with the flux decrease, 
strongly suggesting that the flux change is from the pulsar itself rather 
than another source along the line of sight. 
This is strengthened by the results for higher energy
and narrower time bins (Figure~\ref{fig:2new}),
suggesting that the flux variation occurred within a week or less. 
We also explored 3-day and 1-day binning, but 
count rates are too low to measure when and how quickly the flux 
change occurred. The data hint that it happened within a few days after MJD 55850.

The frequency derivative discontinuity resembles a glitch in $\dot{f}$
\citep[e.g.,][]{cordes1985}. However, the phenomenology differs
from radio and \g-ray glitches \citep[e.g.,][]{espinoza2011,pletsch2012}: 
glitches are not usually associated with a
flux change and are followed by a recovery, not
detected for J2021+4026 prior to MJD~56200.

Doppler shift due to pulsar motion in a binary system cannot
explain the change in $\dot{f}$.
If we assume the pulsar moved in the same direction for the $\sim
3$ years before the jump,
i.e., half of a circular orbit with radius 6 (1) A.U,
that would yield a fractional frequency change due to the Doppler
shift of
$10^{-5}$ ($10^{-4}$), compared to the observed $\sim 4\%$ variation. 
Reproducing the observed change for a 6-year orbit requires a highly eccentric orbit with an
unrealistically small minor axis of 0.01~A.U. Therefore, the
$\dot{f}$ change is likely related to some phenomenon in the
pulsar
magnetosphere.

The jump causes phase coherence loss.
We therefore built two timing solutions using LAT \g~rays
\citep{ray2011}. We used 32-day
intervals to determine pulse times-of-arrival (TOAs)\footnote{ This
yields an integer number of TOAs with
reasonable
pulse profiles.}.
We obtained 36 (13) before (after) the jump.
We used the TEMPO2 package \citep{hobbs2006} to fit these TOAs
using a
model with absolute phase, frequency and its first three derivatives at the reference epoch. 
The RMS of the timing residuals of the post-jump timing solution is 2.1~ms.
The pre-jump solution needed whitening with sinusoidal waves to
achieve a 3.0~ms residual RMS. 
We verified that this is due only to the different lengths of the time ranges.
The timing
solutions\footnote{Available at \url{
http://fermi.gsfc.nasa.gov/ssc/data/access/lat/ephems/ } } confirm
the sudden $\dot{f}$ change near MJD~55850 (Figures~\ref{fig:1} and~\ref{fig:2new}).
Owing to similarities with Geminga, we shifted photon phases
to center the second highest peak at $0.1$, resulting in a half-period
shift relative to 2PC.

\section{Pulsar properties before and after the jump}\label{sec:psrcar}

\begin{figure*}
\epsscale{2.4}
\plottwo{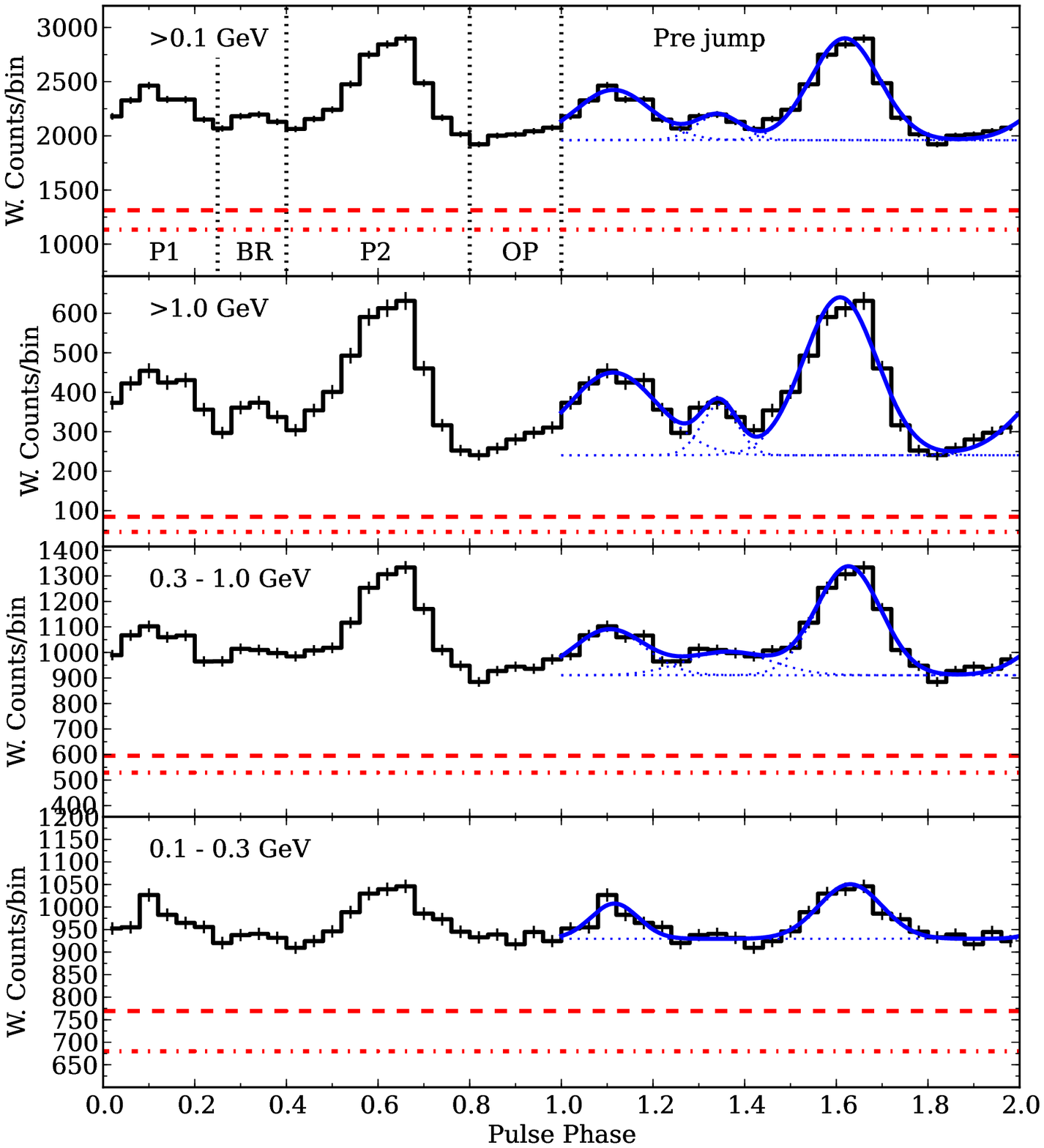}{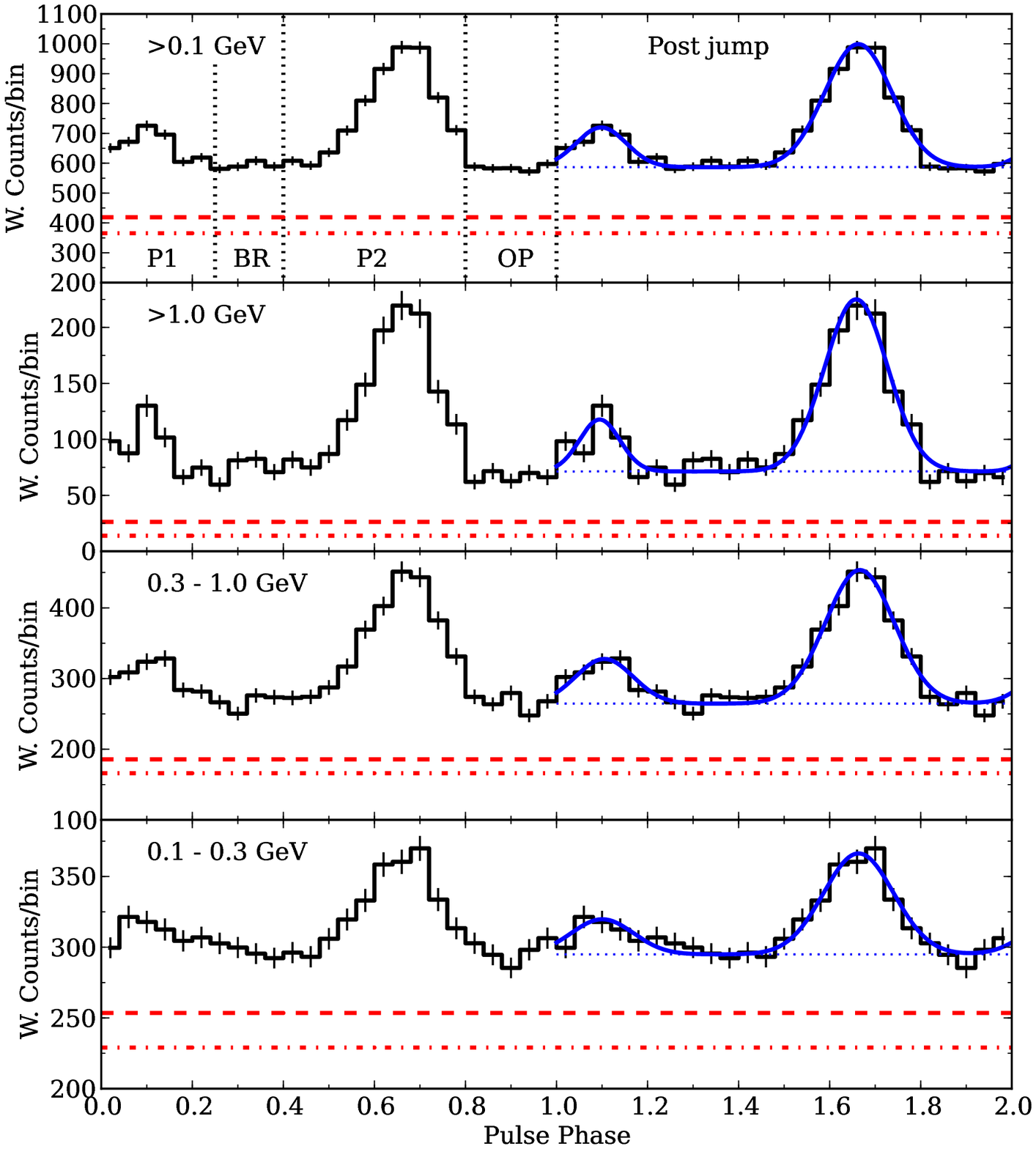}
\caption{Weighted pulse profiles for J2021+4026 in different energy bands,
before (left, 1167 days) and after the jump (right, 423
days). Statistical uncertainties only. 
Red dashed/dash-dotted line: background level from the spectral
fits, including all sources except the pulsar
with/without SNR~G78.2+2.1. 
Fit curves overlay the second rotation: 
blue-dotted show the constant and Gaussian components, 
solid blue show the sums.}\label{fig:2}
\end{figure*}

We studied the pulse profile and the spectrum before and after MJD~55850,
repeating the
likelihood analysis of Section~\ref{sec:anal} for the two time intervals
independently.
Then we selected \g~rays within 2\arcdeg~of the pulsar and used the
best-fit spectrum to assign each photon a weight, the probability of being
associated with
the pulsar.
We assigned each photon a phase using the timing solutions described in Section~\ref{sec:timing}, 
and thus built weighted pulse profiles for different energy bands (Figure~\ref{fig:2}).
The profiles show two main peaks, P1 and P2, interconnected by a bridge (BR),
where a third peak appears before the jump, especially at $E>1$~GeV. 
An off-peak region, OP, follows P2.

We fit the pulse profile peaks with three Gaussians, adding a constant to
account for steady emission (Figure~\ref{fig:2}). 
The third peak in BR is included only when detected with $> 3
\sigma$ significance. 
Table~\ref{tbl-2} summarizes the fit results and other pulsar characteristics.
Across the jump the constant component decreases compared to P2's amplitude, 
and the P1--P2 lag ($\Delta_{12}$) increases. 
The peak height ratio P1/P2 shows hints of a decrease,
while the third peak significance decreases\footnote{The decrease is
partially due to the different epoch lengths before and after
the jump. For the 462 days pre-jump, the third peak detection
significance decreases to $4.6\sigma$ ($2.9\sigma$) for $>100$~MeV ($>1$~GeV).} 
from $8\sigma$ to $2.8\sigma$ ($5.5\sigma$ to $2.2\sigma$) for $>100$~MeV ($>1$~GeV).

To determine the pulsar spectral energy distribution we subdivided
the data before and after the jump into
four phase intervals: $0-0.25$ (P1), $0.25-0.4$ (BR), $0.4-0.8$ (P2), and
$0.8-1$ (OP). For each, we determined the pulsar spectrum over 10 logarithmically-distributed energy
bins from 100
MeV to 10 GeV, approximating the spectrum within each bin with a power law with
spectral index 2. We also determined the
spectral energy distribution over the entire energy band using a PLEC1 model.

\begin{figure*}
\epsscale{2.}
\plotone{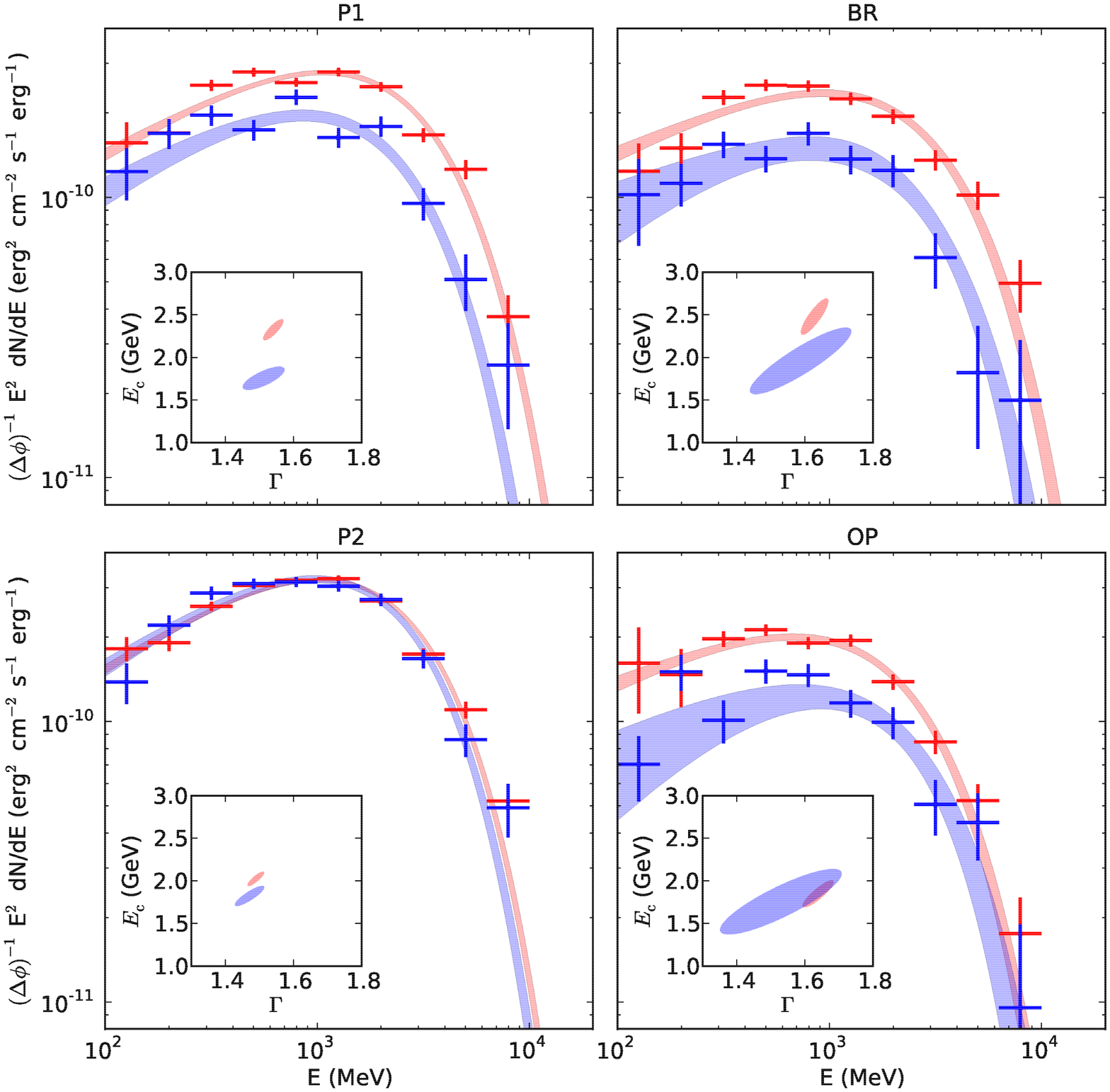}
\caption{Spectral energy distribution of J2021+4026 in four
phase intervals (see text). Spectra are shown for
the time intervals before (red) and after (blue) the jump in the form of flux
points and 1-$\sigma$ contours from the PLEC1 fits (shaded bands). The inset
panels show the covariance ellipses of the spectral index $\Gamma$
and cutoff energy $E_{c}$ for the best-fit PLEC1 model.
Statistical uncertainties only.\label{fig:3}}
\end{figure*}
As shown in Figure~\ref{fig:3}, across the jump the flux varies at all phases but P2, 
strengthening the association of the flux drop with the pulsar as opposed to
another source. The OP spectrum is always well-described by a power law with an
exponential cutoff at $\sim 2$~GeV, 
indicating a magnetospheric origin over all phases.
There is an indication of a
decrease in $E_\mathrm{c}$ for P1 ($\sim 2 \sigma$). 

\section{Summary and Discussion}

We detected a ``jump'', a sudden
decrease of J2021+4026's flux above 100~MeV of $\sim 20\%$
associated with a $\sim 4\%$
increase in spindown rate on a time scale shorter than 1 week. 
The jump is also accompanied by changes in the pulse profile. 
Furthermore, we found evidence for a small, steady flux increase preceding the jump. 
The temporal correlation between spindown and flux changes strongly indicates that these phenomena 
are related to the pulsar. 
While mode changes and other intermittent behavior are well known
for some radio
pulsars \citep[e.g.][]{lyne2010}, this is the first time such behavior has been seen at \g-ray energies. 

J2021+4026 belongs to a small
set of unusual LAT pulsars -- PSR J0633+1746 (Geminga), J1836+5925 and
J2021+4026 -- the sources in 2PC with the brightest
magnetospheric emission at all spin phases.
They are all radio-quiet, with phase lags between the main
peaks $\Delta> 0.5$, higher than typical \citep{2PCpaper}.
Finally, although only Geminga has a parallax
distance (we rely on the SNR association of J2021+4026 and X-ray
spectral arguments for J1836+5925), if we adopt the common assumption that
the \g-ray pulse is effectively uniform on the sky, beaming factor
$f_\Omega=1$, then all three have large efficiencies $\eta = 4\pi f_\Omega F_\gamma
d^2/ \dot{E}_\mathrm{SD} \ge 1$ ($d$ is the distance, $F_{\gamma}$ is the energy
flux, see
Table~\ref{tbl-2}). J2021+4026 is the most extreme of the
three, with
$\eta
= 2.3$.

All of these attributes point to peculiarity in the \g-ray
beaming. They are most easily understood in the context of the classical
outer gap (OG)
model. \citet{romani2010}
show that such large peak lag implies small magnetic
inclinations $\alpha < 30\arcdeg$
and near-equatorial viewing angles $80^\circ < \zeta < 100\arcdeg$. For this
geometry
the pulsars should be radio-quiet, the OG geometry predicts $f_\Omega \approx
0.1-0.2$ ($\eta<1$) and the Earth line-of-sight skims nearly tangentially
to
the peak caustics, producing complex peak structure and strong off-peak emission
\citep{romani2010}.
Also, two-pole caustic (TPC) models \citep{dyks2003} can produce
strong off-peak components for a wider range of geometries (most
with $\alpha < 30^\circ$). These models tend to have single broad
pulses at
small $\zeta$, but large $\zeta$ models can be double pulsed. Thus the
preferred
geometry is similar to that of the OG case, and should also be radio-quiet.
These models have $f_\Omega \approx 0.5-0.75$, making it harder to accommodate
the observed \g-ray flux. If classical TPC solutions are extended to higher
altitude, then one may recover the broad equatorial pulses and small
$f_\Omega$ \citep{pierbattista2013}. The nearly
aligned rotator viewed at high inclination scenario is independently confirmed
for Geminga thanks to X-ray observations of its rotating hot spot
\citep{caraveo2004}.

 When emission from near the light cylinder dominates the pulse, 
the concentration of the 
$\gamma$-ray beam to a narrow
equatorial strip gives high apparent $\eta$ and allows small changes
in magnetic field morphology or even in $\alpha$ to
move a substantial fraction of the $\gamma$-ray beam
over the line of sight, giving
large fractional changes to the pulse profile and $f_\Omega$.
For young pulsars, we expect the \g-ray luminosity to scale
with $\sqrt{\dot{E}_\mathrm{SD}}$ \citep[e.g.,][]{harding1981,2PCpaper}.
The decrease in
flux rate associated
to an increase in spindown rate after the jump strengthens the case
that beaming must play a key role.

Therefore, we can speculate that the
jump of J2021+4026 represents a shift in the magnetic field
structure,
leading
to either an effective $\alpha$ change or an effective current change. 
These may be precipitated by a reconfiguration of field line footpoints at the
surface, i.e. in the crustal layers, that modifies the overall magnetic dipole
torque on the star.
There is
no reason to expect that the resulting spindown increase should enhance the
solid-angle integrated luminosity of the
pulsar $\gamma$-ray emission, since the principal effect is that of a modified
beaming.
If the slow variation in the pulsar flux before the jump
is substantiated by additional study, this might plausibly be associated with
more gradual changes in geometry, for example from force-free precession
\citep[e.g.][]{jones2012}.

The very sensitivity of the beaming to currents and
geometry for the equatorial, small $\alpha$, OG or TPC models
complicates the interpretation of the observations in terms of magnetosphere
configurations.
Alternative tests of these scenarios may rely
on non \g-ray constraints on spin geometry, e.g., from X-ray imaging of the
synchrotron termination shock \citep{ng2004},
or, if radio-loud examples can be detected, from polarization studies.

Radio and X-ray observations have shown that mode changes, and
variability in general, are
key to understanding pulsars \citep[][]{lyne2010,hermsen2013} and
therefore to investigating their fundamental physics
\citep[][]{alpar1984,cordes2004}. The ``jump'' in
J2021+4026 breaks the axiom of pulsars as steady \g-ray emitters, opening new
avenues for pulsar variability studies at \g-ray energies, where the bulk of
their spindown energy is emitted.

\acknowledgments
The \F~LAT Collaboration acknowledges support from a number of agencies and
institutes for both development and the operation of the LAT as well as
scientific data analysis. These include NASA and DOE in the United States,
CEA/Irfu and IN2P3/CNRS in France, ASI and INFN in Italy, MEXT, KEK, and JAXA in
Japan, and the K.~A.~Wallenberg Foundation, the Swedish Research Council and the
National Space Board in Sweden. Additional support from INAF in Italy and CNES
in France for science analysis during the operations phase is also gratefully
acknowledged.

\bibliographystyle{apj}

\end{document}